\documentclass{article}
\usepackage{spconf,amsmath,graphicx,cite}
\usepackage[subtle]{savetrees}
\title{Improving auditory attention decoding performance \\ of linear and non-linear methods using state-space model}
\name{Ali Aroudi \qquad Tobias de Taillez \qquad Simon Doclo}
\address{Department of Medical Physics and Acoustics and Cluster of Excellence Hearing4All,\\ University of Oldenburg, Oldenburg, Germany \\{\tt ali.aroudi@uni-oldenburg.de}
\thanks{This work was supported by the Deutsche Forschungsgemeinschaft (DFG, German Research Foundation) - Project ID 390895286 EXC 2177/1.}}
\begin{document}
\ninept
\maketitle
\begin{abstract}
Identifying the target speaker in hearing aid applications is crucial to improve speech understanding. Recent advances in  electroencephalography (EEG) have shown that it is possible to identify the target speaker from single-trial EEG recordings using auditory attention decoding (AAD) methods. AAD methods reconstruct the attended speech envelope from EEG recordings, based on a linear least-squares cost function or non-linear neural networks, and then directly compare the reconstructed envelope with the speech envelopes of speakers to identify the attended speaker using Pearson correlation coefficients. Since these correlation coefficients are highly fluctuating, for a reliable decoding a large correlation window is used, which causes a large processing delay. In this paper, we investigate a state-space model using correlation coefficients obtained with a small correlation window to improve the decoding performance of the linear and the non-linear AAD methods. The experimental results show that the state-space model significantly improves the decoding performance.
\end{abstract}
%%%%%%%%%%%

%---section
\section{Introduction}
\label{sec:Introduction}
Multi-microphone speech enhancement algorithms in currently available hearing aid devices are able to perform source separation and reduce background noise to improve speech intelligibility. However, the performance of these algorithms in improving speech intelligibility typically depends on correctly identifying the target speaker to be enhanced. In hearing aid applications, the target speaker is typically identified using assumptions such as the target speaker being
located in front of the listener or being the loudest speaker. %Tobi: second "However" in a short time, maybe replace
However, since in real-world conditions these assumptions may often be violated, e.g., when the auditory attention of the listener is misaligned with the assumptions, the performance of speech enhancement methods decreases and results in a substantially reduced benefit from hearing aids.

Recent advances in electroencephalography (EEG) have shown that it is possible to identify the target speaker from single-trial EEG recordings of a listener by decoding the auditory attention \cite{OSullivan_neural_2017, Simon_Eyndhoven_2016, Aroudi_2020_cognitive-driven_beamforming_ASLP}. 
Several auditory attention decoding (AAD) methods have been
proposed to identify the target speaker, based on, e.g., a linear least-squares cost function \cite{OSullivan2014, Wong_Frontiers_2018, Alickovic_2019, Sina_2018} and non-linear neural networks \cite{Tobias_2018, Gregory_2019}.
The linear least-squares-based AAD method proposed in \cite{OSullivan2014} is able to exploit the linear neural process of attention along the auditory pathway to identify the attended speaker. The non-linear neural-network-based AAD method proposed in \cite{Tobias_2018} is able to exploit the non-linear neural process of attention in addition to the linear neural process. To identify the attended speaker, these methods aim at reconstructing the attended speech envelope from the EEG recordings using a trained spatio-temporal estimator. 
In the training step, a spatio-temporal envelope estimator is trained by either minimizing the least-squares error or maximizing the correlation cost function between the attended speech envelope and the reconstructed envelope. 
In the decoding step, the attended speech envelope is reconstructed using the trained envelope estimator and then directly compared with the speech envelopes of two speakers using Pearson correlation coefficients to identify the attended speaker. 
Since these correlation coefficients are highly fluctuating,  %long EEG trials 
for a reliable decoding a large correlation window on the order of $30$ seconds is typically used, which causes a large processing delay and hence limits the feasibility of AAD for hearing aid applications. 
In \cite{Sina_2018}, it has been proposed to 
use coefficients of least-squares-based envelope estimators, obtained separately for reconstructing the attended and the unattended speech envelope.
Using coefficients of estimators, a state-space model then identifies the attended speaker. 
%employ coefficients of least-squares-based estimators, obtained separately for reconstructing the attended and unattended envelopes, instead of correlation coefficients to identify the attended speaker. 
%Using a state-space model, the estimator coefficients are then translated into probabilistic measures of the attention state, from which the attended speaker is identified. 
%Since obtaining reliable estimator coefficients depends on signal-to-noise ratio of EEG signals, a batch-mode analysis and an EEG channel selection are performed prior to AAD, which may not be desired for real-time applications.
%not be desired for real-time applications.  
%Tobi: what about:
% In this paper, we investigate how an state-space model using correlation coefficients obtained with a small correlation windows can improve this processing delay to identify the attended speaker more quickly. 
In this paper, we investigate a state-space model using correlation coefficients obtained with a small correlation window to improve the decoding performance of the  (linear) least-squares-based AAD method and the (non-linear) neural-network-based AAD method. The correlation coefficients are generated using either the least-squares-based AAD method or the neural-network-based AAD method. The state-space model then translates the generated correlation coefficients into smooth estimates of the attention state, based on which the attended speaker is identified.
%identify the attended speaker. The correlation coefficients are generated using either the (linear) least-squares-based AAD method and the (non-linear) neural-network-based AAD method. The state-space model then translates the generated correlation coefficients into smooth estimates of the attention state, based on which the attended speaker is identified.

%to improve the decoding performance of the (linear) least-squares-based AAD method or the (non-linear) neural-network-based AAD method. The least-squares-based method and the neural-network-based AAD method generate correlation coefficients using a $5$-second correlation window. The state-space model then translates the fluctuating correlation coefficients into smooth estimates of the attention state.

For an acoustic scenario with two competing speakers and diffuse noise at different SNRs and reverberation times, 64-channel EEG responses with 18 participants were
recorded. The experimental results show for correlation coefficients obtained with a $5$-second correlation window that the least-squares-based AAD method and the neural-network-based AAD method yield a low decoding performance. However, when using the state-space model with the least-squares-based AAD method, the decoding performance  significantly improves.
%between using the least-squares-based AAD method and the neural-network-based AAD method. However, when using the state-space model with the least-squares-based AAD method, the decoding performance  significantly improves.

%the state-space model improves the decoding performance, in particular when using with the least-squares-based AAD method. In addition, the results showed that the least-squares-based AAD method and the neural-network-based AAD method yield a very similar decoding performance.

%---section
\section{Auditory attention decoding}
\label{sec:Auditory attention decoding}
This section presents the auditory attention decoding using a state-space model, which   employs correlation coefficients generated either by the least-squares-based AAD method and the neural-network-based AAD method. In Section \ref{subsec:Configuration and notation} the acoustic scenario and the notation are defined. Section \ref{subsec: AAD using state-space model} describes the state-space model. Section \ref{subsec: Least-squares-based AAD} and  Section \ref{subsec: Neural-network-based AAD} describe the least-squares-based AAD method and the neural-network-based AAD method.
%---subsection
\subsection{Configuration and notation}
\label{subsec:Configuration and notation}
\begin{figure}[t]
 \centering
  \centerline{\includegraphics[width=6cm,height=8cm]{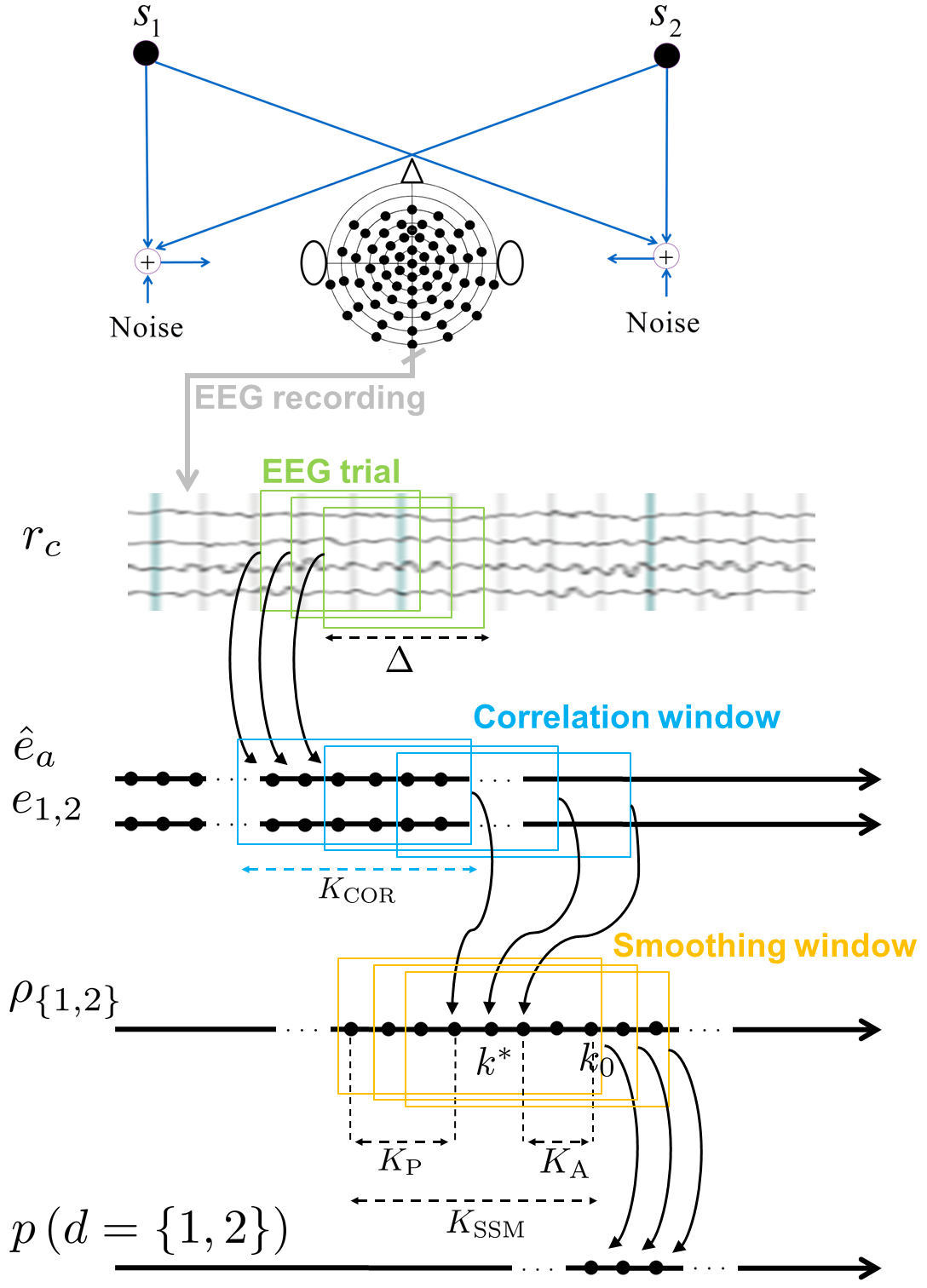}}
 \caption{Illustration of the process flow of AAD using state-space model.}
\label{fig:decoding flow process}
\end{figure}

We consider an acoustic scenario comprising two competing speakers and background noise in a reverberant environment, where the ongoing EEG responses of a listener to these acoustic stimuli are recorded (See Fig. \ref{fig:decoding flow process}). The clean speech signal of speaker $1$ is denoted as $s_{1}\left[n\right]$, with $n$ the discrete time index, while the clean speech signal of speaker $2$ is denoted as $s_{2}\left[n\right]$. The envelopes of the clean speech signals of speaker $1$ and $2$ are denoted as  $e_{1}\left[k\right]$ and $e_{2}\left[k\right]$, with $k$ the sub-sampled time index, respectively. 

%The EEG recordings are segmented into trials. 
The reconstructed attended speech envelope from $C$-channel EEG recordings $r_{c}\left[k\right]$, with $c=1\ldots C$, using a trained spatio-temporal envelope estimator $F$ is given by 
%When using a trained (linear or nonlinear) spatio-temporal estimator $F$ to reconstruct the speech envelope of the attended speaker from $C$-channel EEG recordings, with $r_{c}\left[k\right]$, $c=1\ldots C$, the reconstructed envelope is given by
%The speech envelope of the attended speaker $\hat{e}_{a}\left[k\right]$ is reconstructed from $C$-channel EEG recordings, with $r_{c}\left[k\right]$, $c=1\ldots C$, using a trained (linear or nonlinear) spatio-temporal estimator $F$, i.e., 
\begin{equation}
\hat{e}_{a}\left[k\right]=F(\mathbf{r}\left[k\right]),   \end{equation}
with
\begin{equation}
\mathbf{r}\left[k\right]=\left[\mathbf{r}_{1}^{T}\left[k\right]\,\mathbf{r}_{2}^{T}\left[k\right]\;\ldots\;\mathbf{r}_{C}^{T}\left[k\right]\right]^{T},
\end{equation}
\begin{equation}
\mathbf{r}_{c}\left[k\right]=\left[r_{c}\left[k\right]\,r_{c}\left[k+1\right]\;\ldots\;r_{c}\left[k+\Delta\right]\right]^{T},
\label{eq: eeg of a channel}
\end{equation}
where $\Delta$ denotes the latency considered for modeling the attentional effect in the EEG responses to acoustic stimuli.

The Pearson correlation coefficients between the reconstructed attended envelope $\hat{e}_{a}\left[k\right]$ and the envelope of two speakers are given by
\begin{equation}
\rho_{1,k}=\rho\left(\mathbf{e}_{1}\left[k\right],\;\hat{\mathbf{e}}_{a}\left[k\right]\right),\;\ \rho_{2,k}=\rho\left(\mathbf{e}_{2}\left[k\right],\;\hat{\mathbf{e}}_{a}\left[k\right]\right),
\label{eq:corr coff}
\end{equation}
where $\hat{\mathbf{e}}_{a}\left[k\right]$ denotes the stacked vector of the reconstructed attended envelope corresponding to a correlation window of length $K_{\text{COR}}$, i.e., 
\begin{equation}
%\mathbf{e}_{1}\left[k\right]=\left[e_{1}\left[k-K_{\text{COR}}+1\right]\,e_{1}\left[k-K_{\text{COR}}+2\right]\;\ldots\;e_{1}\left[k\right]\right]^{T},
\hat{\mathbf{e}}_{a}\left[k\right]=\left[\hat{e}_{a}\left[(k-1)K_{\text{COR}}+1\right]\,\hat{e}_{a}\left[(k-1)K_{\text{COR}}+2\right]\;\ldots\;\hat{e}_{a}\left[kK_{\text{COR}}\right]\right]^{T},
\label{eq: stacked vector of the reconstructed envelope}
\end{equation}
and $\mathbf{e}_{1}\left[k\right]$ and $\mathbf{e}_{2}\left[k\right]$ are defined similarly as in (\ref{eq: stacked vector of the reconstructed envelope}). 
Please note that in this paper we assume that the clean speech signal of speakers are available for obtaining the envelopes of speakers $e_{1}\left[k\right]$ and $e_{2}\left[k\right]$. However, since in practice only  microphone signals containing a mixture of speakers and ambient noise are available, the clean speech signal of speakers needs to be appropriately estimated from microphone signals, e.g., by using the noise reduction and source separation algorithms proposed in \cite{Aroudi_2020_cognitive-driven_beamforming_ASLP, Aroudi_ICASSP2019, Cong_2019, Simon_Eyndhoven_2016}. 

%The attention state of the listener when attending to either speaker $1$ or $2$ is defined as a binary random variable following a Bernoulli process, i.e.,
%\begin{equation}
%\left\{ \begin{array}{cc}
%d_{k}=1 & \text{when attending to speaker 1}\\
%d_{k}=0 & \text{when attending to speaker 2}
%\end{array}\right.
%\end{equation}

%---subsection
\subsection{AAD using state-space model}
\label{subsec: AAD using state-space model}
Suppose the attended envelope is reconstructed using a trained (linear or nonlinear) spatio-temporal estimator and the correlation coefficients of speakers are obtained. 
We aim at estimating the probability of attending to speaker $1$ or $2$ based on a state-space model using the past and the subsequent correlation coefficients (see Fig. \ref{fig:decoding flow process}). 
Let the attention state of the listener when attending to either speaker $1$ or $2$ be defined as a binary random variable, i.e.,
\begin{equation}
\left\{ \begin{array}{cc}
d_{k}=1, & \text{when attending to speaker 1}\\
d_{k}=2, & \text{when attending to speaker 2}
\end{array}\right.,
\end{equation}
which follows a Bernoulli process. 
The probability of attending to speakers based on the state-space model is obtained as \cite{Sahar_2016, Sina_2018}
\begin{equation}
p\left(d_{k}=1\right)=1-p\left(d_{k}=2\right)=\dfrac{1}{1+e^{-\left(z_{k}\right)}},
\label{eq: prob_1_and_2}
\end{equation}
with
\begin{equation}
z_{k}=c_{0}z_{k-1}+w_{k},
\label{eq: autoregressive}
\end{equation}
\begin{equation}
w_{k}\sim\mathcal{N}\left(0,\eta_{k}\right),
\end{equation}
\begin{equation}
\eta_{k}\sim\textrm{Inverse-Gamma}\left(a_{0},b_{0}\right),
\label{eq: Inverse-Gamma}
\end{equation}
, $c_{0}$ denoting the hyperparameter ensuring the stability of  $z_{k}$, and  $a_{0}$ and $b_{0}$ denoting the hyperparameters used to control the smoothing degree of the state-space model by tuning the variations of $z_{k}$ and $p\left(d_{k}=\left\{ 1,2\right\} \right)$.
The autoregressive model in (\ref{eq: autoregressive}) implies that the (attention state) parameter $z_{k}$ at instance $k$ is predicted from $z_{k-1}$ at the instance $k-1$ with some uncertainty, which is modeled by the noise process $w\left(k\right)$. Please note that when $z_{k}$ varies from $-\infty$ to $\infty$, $p\left(d_{k}=1\right)$ monotonically varies from $0$ to $1$.
%The hyperparameters $a_{0}$ and $b_{0}$ can be tuned to avoid large variations of $z_{k}$ in (\ref{eq: autoregressive}), and consequently $p\left(d_{k}=1\right)$ in (\ref{eq: prob_1_and_2}). 
To relate the correlation coefficients of speakers to the attention state, the probability of the absolute values of correlation coefficients given attending to either speaker $1$ or $2$ is modeled using a Log-Normal distribution, i.e.,
\begin{equation}
\begin{array}{cc}
p\left(\left|\rho_{l,k}\right|\mid d_{k}=l\right)\sim\textrm{Log-Normal}\left(\boldsymbol{\alpha}_{a}\right), & l=1,2\end{array}
\label{eq: atteneded Log-Normal distribution}
\end{equation}
with $\boldsymbol{\alpha}_{a}$ denoting the parameter set of the attended Log-Normal distribution.
The probability of the correlation coefficients given ignoring either speaker $1$ or $2$ is modeled as 
\begin{equation}
\begin{array}{cc}
p\left(\left|\rho_{l,k}\right|\mid d_{k}\neq l\right)\sim\textrm{Log-Normal}\left(\boldsymbol{\alpha}_{u}\right), & l=1,2\end{array}
\label{eq: unatteneded Log-Normal distribution}
\end{equation}
with $\boldsymbol{\alpha}_{u}$ denoting the parameter set of the unattended Log-Normal distribution. 
%The parameter set of the attended and the unattended Log-Normal distribution can be tuned such that the Log-Normal distributions are separated with minimal probabilistic overlapping.

Let's suppose we are at the instance $k=k_{0}$ (see Fig. \ref{fig:decoding flow process}) and aim to estimate the probability of attending to speakers  $p\left(d_{k}=\left\{ 1,2\right\} \right)$ at the instance $k=k^{\ast}$ using the correlation coefficients obtained within a sliding smoothing window of length $\ensuremath{\ensuremath{K_{\text{SSM}}}}=K_{P}+K_{A}+1$, with $K_{P}$ and $K_{A}$ denoting the parameters determining the number of the correlation coefficient prior to and after the instance $k^{\ast}$, respectively.  The parameters of the state-space model corresponding to the smoothing window are hence given as $\Omega=\left\{ z_{k_{0}-\ensuremath{K_{\text{SSM}}+1}:k_{0}},\eta_{k_{0}-\ensuremath{K_{\text{SSM}}+1}:k_{0}},\boldsymbol{\alpha}_{a},\boldsymbol{\alpha}_{u}\right\} $. Theses parameters including $z_{k^{\ast}}$ are estimated from the correlation coefficients $\rho_{1,k_{0}-\ensuremath{K_{\text{SSM}}+1}:k_{0}}$ and $\rho_{2,k_{0}-\ensuremath{K_{\text{SSM}}+1}:k_{0}}$ obtained within the smoothing window  using the  Expectation Maximization (EM) estimation algorithm proposed in \cite{Sahar_2016, Sina_2018}. Based on the estimated attention state parameter $z_{k^{\ast}}$, the probability of attending to speakers $p\left(d_{k^{\ast}}=\left\{ 1,2\right\} \right)$ are obtained. It is then decided that the listener attended to speaker 1 if $p\left(d_{k^{\ast}}=1\right)>p\left(d_{k^{\ast}}=2\right)$ or attended to speaker 2 otherwise. Please note that the estimated parameters $\Omega$ are also used for the initialization of parameters in the next smoothing window.
%The parameter $K_{F}$ controls the trade-off between real-time estimation and robust estimation of ?, i.e., for $K_{F}=0$ the estimation is performed in real-time while lacking the robustness to the fluctuations of the correlation coefficients. 

In the simulations (see Section \ref{sec:Experimental setup}), we will consider to use the state-space model with correlation coefficients generated either by the least-squares-based AAD method (see Section \ref{subsec: Least-squares-based AAD}) or the neural-network-based AAD method (see Section \ref{subsec: Neural-network-based AAD}).
%decode auditory attention of a listener using the state-space model and the correlation coefficients generated by either the least-squares-based AAD method (see Section \ref{subsec: Least-squares-based AAD}) or the neural-network-based AAD method (see Section \ref{subsec: Neural-network-based AAD}).

%---subsection
\subsection{Least-squares-based AAD}
\label{subsec: Least-squares-based AAD}
The least-squares-based AAD method proposed in \cite{OSullivan2014} aims at estimating the attended speech envelope from the EEG recordings using a trained linear spatio-temporal estimator. 
In the training step, the attended speaker is assumed to be
known and an attended speech signal is used to train a linear estimator by minimizing the least-squares error between the attended speech envelope $e_{a}\left[k\right]$ and the reconstructed envelope $\hat{e}_{a}\left[k\right]$, i.e.,
%\begin{equation}
%\underset{\mathbf{g}}{\textrm{min}}\frac{1}{K}\overset{K}{\underset{k=1%}{\sum}}(e_{a}\left[k\right]-\underset{\hat{e}_{a}\left[k\right]}{\unde%rbrace{\mathbf{g}^{T}\mathbf{r}\left[k\right]}})^{2}+\beta\mathbf{g}^{T%}\mathbf{D}\mathbf{g},
%\label{eq: LS cost function}
%\end{equation}
\begin{equation}
\underset{\mathbf{g}}{\textrm{min}}\frac{1}{K}\overset{K}{\underset{k=1}{\sum}}\left(e_{a}\left[k\right]-\hat{e}_{a}\left[k\right]\right)^{2}+\beta\mathbf{g}^{T}\mathbf{D}\mathbf{g},
\label{eq: LS cost function}
\end{equation}
with $\hat{e}_{a}\left[k\right]=F(\mathbf{r}\left[k\right])=\mathbf{g}^{T}\mathbf{r}\left[k\right]$,
%\begin{equation}
%\hat{e}_{a}\left[k\right]=F(\mathbf{r}\%left[k\right])=\mathbf{g}^{T}\mathbf{r}%\left[k\right],
%\end{equation}
 $\mathbf{D}$ denoting the derivative matrix \cite{Aroudi_EMBC_2017} and $\beta$ denoting a regularization parameter. The linear estimator minimizing the regularized least-squares cost function in (\ref{eq: LS cost function}) is equal to 
\begin{equation}
\mathbf{g}=\left(\mathit{\mathbf{Q}+\beta\mathbf{D}}\right)^{-1}\mathbf{q},
\label{eq:LS estimator}
\end{equation}
with the correlation matrix $\mathbf{Q}$ and the cross-correlation vector $\mathbf{q}$ given by
\begin{equation}
\mathbf{Q}=\mathit{\frac{\mathrm{1}}{K}\overset{K}{\underset{k=\mathrm{1}}{\sum}}\left(\mathbf{r}\left[k\right]\mathbf{r}^{T}\left[k\right]\right)},\;\ \mathbf{q}=\mathit{\frac{\mathrm{1}}{K}\overset{K}{\underset{k=\mathrm{1}}{\sum}}\left(\mathbf{r}\left[k\right]e_{a}\left[k\right]\right)}.
\label{eq: correlatin matrix and vector}
\end{equation}

In the decoding step, the attended envelope $\hat{e}_{a}\left[k\right]$ is obtained using the (trained) linear estimator $\mathbf{g}$ in (\ref{eq:LS estimator}). Next, the correlation coefficients between the reconstructed attended envelope and the envelope of two speaker $\rho_{1,k}$ and $\rho_{2,k}$ are computed as in (\ref{eq:corr coff}).  
Based on these correlation coefficients, it is then decided that the listener attended to speaker 1 if $\rho_{1,k}>\rho_{2,k}$ or attended to speaker 2 otherwise.

%---subsection
\subsection{Neural-network-based AAD}
\label{subsec: Neural-network-based AAD}
The neural-network-based AAD method aims at estimating the attended speech envelope from the EEG recordings using a trained non-linear spatio-temporal estimator. Similarly as in \cite{Tobias_2018, Gregory_2019}, we consider a network $\mathcal{H}$ consisting of a hidden convolutional layer with hyperbolic tangent activation functions and one output layer with linear activation functions. 
In the training step, the network is trained to maximize the correlation between the attended speech envelope and the reconstructed envelope by minimizing the correlation cost function \cite{Tobias_2018}, i.e.,
\begin{equation}
\text{min \ensuremath{\frac{1}{K}\overset{K}{\underset{k=1}{\sum}}}}(1-\rho\left(\mathbf{e}_{a}\left[k\right],\;\hat{\mathbf{e}}_{a}\left[k\right]\right)),
\end{equation}
A correlation cost function equal to $0$ corresponds to the maximum correlation between the attended speech envelope and the reconstructed envelope, i.e., $\rho\left(\mathbf{e}_{a}\left[k\right],\;\hat{\mathbf{e}}_{a}\left[k\right]\right)=1$, while a correlation cost function equal to $1$ corresponds to the minimum correlation. A correlation cost function larger than $1$ corresponds to a negative correlation.
%To avoid over-fitting, the dropout technique in \cite{Srivastava_2014} is used.  

In the decoding step, the attended envelope is obtained using the (trained) network $\mathcal{H}$, i.e., $\hat{e}_{a}\left[k\right]=F(\mathbf{r}\left[k\right])=\mathcal{H}(\mathbf{r}\left[k\right])$.
%\begin{equation}
%\hat{e}_{a}\left[k\right]=F(\mathbf{r}\%left[k\right])=\mathcal{H}(\mathbf{r}\l%eft[k\right]).
%\end{equation} 
Next, the correlation coefficients between the reconstructed attended envelope and the envelope of two speaker are computed  $\rho_{1,k}$ and $\rho_{2,k}$ as in (\ref{eq:corr coff}).  
Based on these correlation coefficients, it is then decided that the listener attended to speaker 1 if $\rho_{1,k}>\rho_{2,k}$ or attended to speaker 2 otherwise.

\section{Experimental setup}
\label{sec:Experimental setup}

%---subsection
\subsection{Acoustic stimuli and EEG measurement}
EEG responses were recorded for $18$ native German-speaking participants. Two German audio stories, uttered by two different male speakers, were simultaneously presented to the participants using 
%Tobi: it is "in ear headphones", isn't it? 
insert earphones. The presented stimuli at both ears were generated by convolving the clean speech signals, i.e., the audio stories, with (non-individualized) binaural impulse responses from \cite{Kayser2009}, and adding diffuse noise, generated according to \cite{Habets2008}. The left and the right speaker were simulated at $\theta_{1}=-45^{\circ}$ and $\theta_{2}=45^{\circ}$. 
Eight different acoustic conditions were considered for the stimuli: one anechoic condition with no background noise, two reverberant conditions with a moderate and a large reverberation time (reverberation time $T_{60}=0.5$ s and $1$ s), two anechoic conditions with %Tobi: I expected "binaural noise" and not "binaural input" here
binaural input $\mathrm{SNRs}=9.0$ dB and $4.0$ dB, and three combinations of reverberation and noise. Among all participants, $8$ participants were instructed to attend to the left speaker, while $10$ participants were instructed to attend to the right speaker. Two participants were excluded from the analysis, one participant due to poor attentional performance and the other one due to a technical hardware problem. 
%Tobi: Leaving you with how many left/right attendees? 6/10 oder 8/8 ?
The EEG responses were recorded using $C=64$ channels at a sampling frequency of $500$ Hz, and referenced to the nose electrode. The EEG responses were re-referenced offline to a common average reference, band-pass filtered between $2$ Hz and $8$ Hz using a third-order Butterworth band-pass filter, and subsequently downsampled to $64$ Hz. The envelopes of the speech signals were obtained using a Hilbert transform, followed by low-pass filtering at $8$ Hz and downsampling to $64$ Hz. 

%---subsection
\subsection{AAD training and testing}
For AAD training and testing, the EEG recordings for the different acoustic conditions were grouped together based on acoustic similarity, resulting in four experimental analysis conditions, i.e., anechoic, reverberant, anechoic-noisy, and reverberant-noisy, each of length $20$ minutes. 
To avoid using EEG recordings of the same experimental analysis condition for training and testing, the leave-one-condition-out approach was used, i.e., four combinations of three experimental analysis conditions without repetition were considered for training and the left condition for each combination was considered for testing. This resulted in four training conditions and four testing conditions. 
%resulting in four training conditions comprising EEG recordings of combinations of three experimental analysis conditions, and four testing datasets comprising EEG recordings of the remaining condition for each combination.
%resulting in four training datasets comprising   EEG recordings of combinations of three experimental analysis conditions, as well as four testing datasets comprising EEG recordings of the remaining condition.
%we considered the training datasets consisting EEG recordings of combinations of three experimental analysis conditions and testing datasets of the remaining 
%the EEG recordings of each three conditions were assigned to a training dataset and the EEG recordings of the remaining condition were assigned to a testing dataset, resulting in four different training and testing datasets. 
%the leave-one-condition-out cross validation approach was used, i.e, when testing using EEG recordings of, e.g., the reverberant-noisy condition, we performed training using EEG recordings of the anechoic, reverberant and noisy conditions, we performed testing using EEG recordings of the reverberant-noisy condition. 
%For training, k-fold cross-validation approach with $k=10$, each of length $6$ minutes, was used to find the optimal parameters of estimators. 
%For training, the first $54$ minutes (corresponding to $90\%$) of each training dataset were used for training estimators and the remaining $6$ minutes (corresponding to $10\%$) were used for evaluation. 

For the least-squares-based AAD method, the latency parameter of the linear estimator in (\ref{eq: eeg of a channel}) was set to $\Delta=20$ (corresponding to $312$ ms), as found to be an appropriate choice for AAD \cite{OSullivan2014, Ali_2019_neural_system}.
%to ensure capturing the attentional effect in the EEG responses. 
For training, the estimator in (\ref{eq:LS estimator}) and the regularization parameter $\beta$ of the estimator in (\ref{eq: LS cost function}) was determined using a k-fold cross-validation approach with $k=10$, each of length $6$ minutes. For testing, the EEG recordings were segmented into trials of length $5$ s with an overlap of $4.98$ s (corresponding to one sample shift). The correlation coefficients were computed using a correlation window of length $\ensuremath{K_{\text{COR}}}=5$ s with an overlap of $4.5$ s. 

For the neural-network-based AAD method, the network $\mathcal{H}$ consisting of a convolutional hidden layer with a filter kernel size of 20 samples (corresponding to $312$ ms) was used. 
For training, the network was trained using a k-fold cross-validation approach with $k=10$, each of length $6$ minutes. The network was trained with the Nadam optimizer \cite{Dozat2016} using a batch size of 3840 samples (corresponding to $60$ seconds $\times$ $64$ channels), a learning rate of $0.002$, and $3000$ iterations.  To avoid over-fitting, the dropout technique from \cite{Srivastava_2014} was used with a ratio of $0.25$, which corresponds to randomly setting $25\%$ of the hidden units to $0$. 
%i.e., $25\%$ of the hidden units was randomly set to $0$. 
The network was implemented in Keras \cite{chollet2015keras}. For testing, the correlation coefficients were obtained using the same correlation window setting as used for the least-squares-based AAD method. 

For the state space model, the hyperparameters $c_{0}$ in (\ref{eq: autoregressive}) and $a_{0}$ and $b_{0}$ in (\ref{eq: Inverse-Gamma}) were set to $c_{0}=1$, $a_{0}=2.008$ and $b_{0}=0.2016$, similarly as in \cite{Sina_2018}. For testing, a sliding smoothing window of length $\ensuremath{K_{\text{SSM}}}=3$ with $K_{A}=1$, $K_{P}=1$ was used. 
For each testing condition, the parameter set of the attended Log-Normal distribution $\boldsymbol{\alpha}_{a}$ in (\ref{eq: atteneded Log-Normal distribution}) was initialized by fitting over correlation coefficients of the (oracle) attended speaker obtained during the first $15$ s and was then fixed. The parameter set of the unattended Log-Normal distribution $\boldsymbol{\alpha}_{u}$ in (\ref{eq: unatteneded Log-Normal distribution}) was similarly initialized by fitting over correlation coefficients of the (oracle) unattended speaker. For testing, the parameters of the state-space model $\Omega$ corresponding to an smoothing window were estimated using the EM estimation algorithm with $20$ iterations.

%To estimate the parameters of the state-space model $\Omega$ corresponding to an smoothing window, the EM estimation algorithm with $20$ iterations was used.

The quality of correlation coefficients generated by either the least-squares-based AAD method or the neural-network-based AAD method was evaluated in terms of the attended correlation and the unattended correlation. The attended correlation was computed using the Pearson correlation between the reconstructed envelopes and the envelopes of the attended speaker, i.e., $\rho_{a,k}=\rho\left(\mathbf{e}_{a}\left[k\right],\;\hat{\mathbf{e}}_{a}\left[k\right]\right)$. The unattended correlation was computed between the reconstructed envelopes and the envelopes of the unattended speaker, i.e. $\rho_{u,k}=\rho\left(\mathbf{e}_{u}\left[k\right],\;\hat{\mathbf{e}}_{a}\left[k\right]\right)$, 
%and the unattended correlation were computed using the Pearson correlation between the reconstructed envelopes and the envelopes of the attended speaker and the unattended speaker, respectively.

The decoding performance was evaluated for several AAD methods, i.e., the least-squares-based AAD method (LS), the neural-network-based AAD method (NN), the state-space model using with the least-squares-based AAD method (LS-SSM) and the state-space model with the neural-network-based AAD method (NN-SSM). 
%The decoding performance of the least-squares-based AAD method, the neural-network-based AAD method, the state-space model using with the least-squares-based AAD method and the state-space model with the neural-network-based AAD method were evaluated. 
The decoding performance for the least-squares-based and the neural-network-based AAD method was computed as the percentage of correctly decoded $5$-second correlation windows. The decoding performance for the state-space model using either the least-squares-based or the neural-network-based AAD method was computed as the percentage of correctly decoded $5$-second smoothing windows.
%The decoding performance was evaluated using the following AAD methods:
%\begin{itemize}
%    \item LS: least-squares-based AAD method
%    \item NN: neural-network-based AAD method 
%    \item LS--SSM: state-space model with the least-squares-based AAD method
%    \item NN--SSM: state-space model with the neural-network-based AAD method 
%\end{itemize}

%When using the state-space model for AAD (LS-SSM algorithm and the NN-SSM algorithm), the decoding performance was computed as the percentage of correctly decoded $5$-second smoothing windows. 

%==Figure
\begin{figure}[t]
\begin{minipage}[b]{.50\linewidth}
  \centering
  \centerline{\includegraphics[width=4.7cm]{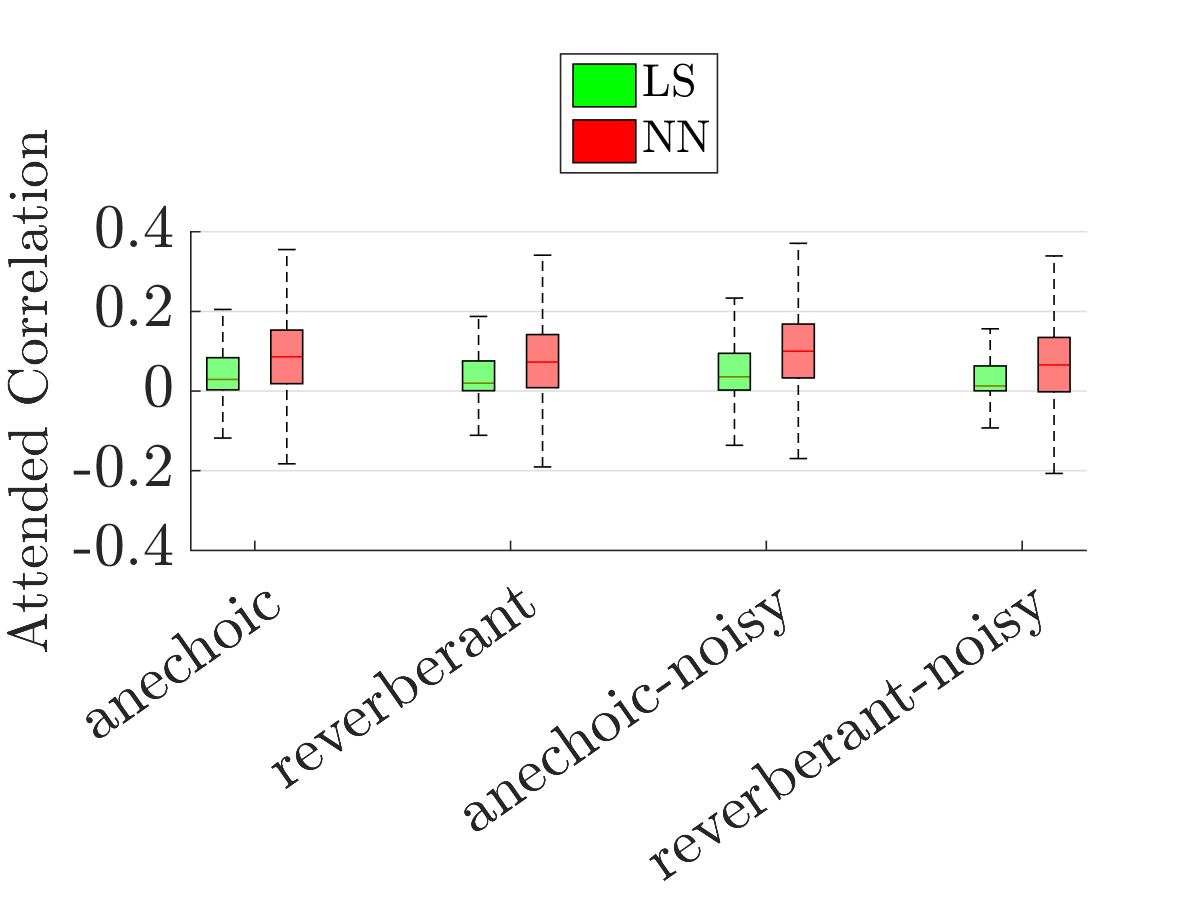}}
%  \vspace{1.5cm}
%\centerline{\;\;\;\;\;\;\;\;\;\;\;\;(a) anechoic-noisy condition}\medskip
\end{minipage}%
\begin{minipage}[b]{.55\linewidth}
  \centering
  \centerline{\includegraphics[width=4.7cm]{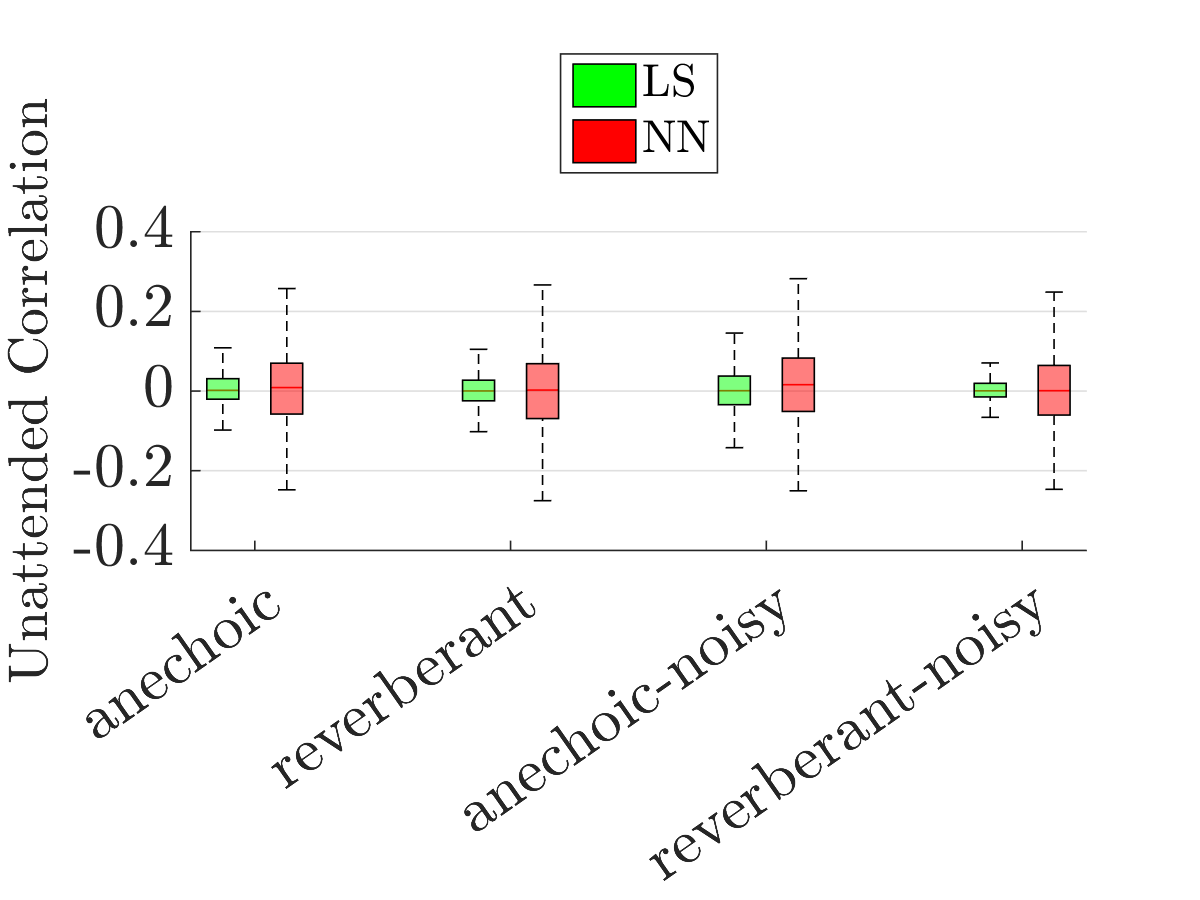}}
%  \vspace{1.5cm}
%\centerline{\;\;\;\;\;\;\;\;\;\;\;\;(b) reverberant-noisy condition}\medskip
\end{minipage}
\caption{Attended correlation and unattended correlation for different acoustic conditions when using the least-squares-based method and the neural-network-based AAD method.}
\label{fig: attended and unattended correlations}
\end{figure}

\section{Results and discussion}
\label{sec:Experimental Results}
%==Figure
\begin{figure}[t]
  \centering
  \centerline{\includegraphics[width=7cm]{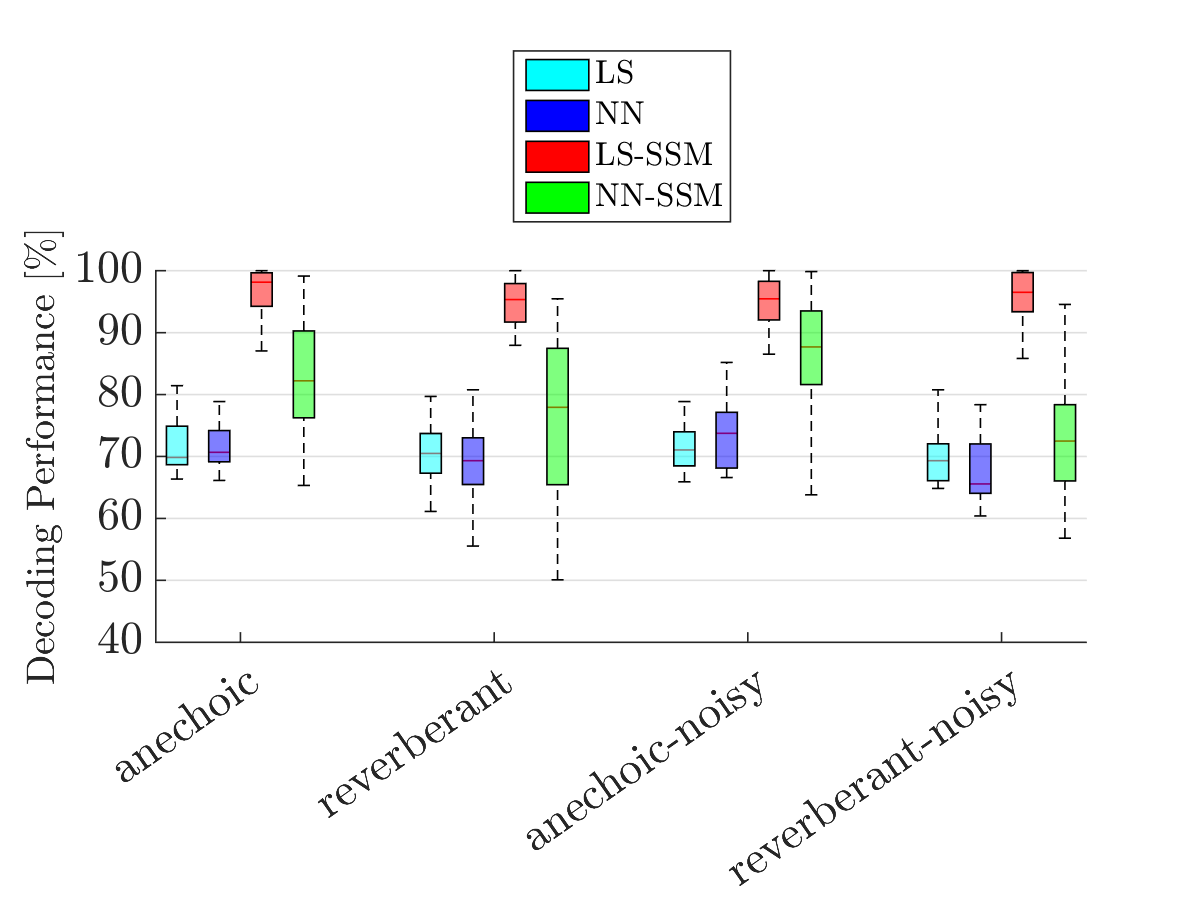}}
%  \vspace{1.5cm}
%\centerline{\;\;\;\;\;\;\;\;\;\;\;\;(a) anechoic-noisy condition}\medskip
%
\caption{Decoding performance for different acoustic conditions when using the least-squares-based method, the neural-network-based AAD method, the state-space model with the least-squares-based method and the state-space model with the neural-network-based AAD method.}
\label{fig: decoding performance}
\end{figure}

%\begin{figure}[t]
%  \centering
%  \centerline{\includegraphics[width=8cm]{./fig/cc/_K_F1_K_B1_fixed_corr_}}
%%  \vspace{1.5cm}
%\centerline{\;\;\;\;\;\;\;\;\;\;\;\;(a) anechoic-noisy condition}\medskip
%%
%\caption{Average decoding performance and binaural $\textrm{SINR}$ improvement ($\Delta\textrm{SINR}$) for the anechoic-noisy and reverberant-noisy conditions. The red dashed-line represents the upper boundary of the confidence interval corresponding to chance level based on a binomial test at the $5$\% significance level. The error bars represent the $95$\% bootstrap confidence interval.}
%\label{fig: Performance}
%%
%\end{figure}

For the least-squares-based AAD method and the neural-network-based AAD method, Fig. \ref{fig: attended and unattended correlations} depicts the attended correlation and the unattended correlation for different acoustic conditions. 
It can be observed for all acoustic conditions that 
the attended correlation obtained by the neural-network-based AAD method (NN) is larger than the least-squares-based AAD method (LS), showing that the neural-network-based AAD method is able to reconstruct the attended speech envelope with a better accuracy. However, the attended correlation obtained by the neural-network-based AAD has a larger variability compared to the least-squares-based AAD method, which corresponds to attended correlation coefficients with a larger fluctuation. In addition, it can be observed that there is no significant difference in the unattended correlation obtained by the least-squares-based AAD method and the neural-network-based AAD method. However, the unattended correlation obtained by the neural-network-based AAD has a larger variability compared to the least-squares-based AAD method.

%the attended correlation obtained by the neural-network-based AAD method is significantly larger than using the least-squares-based AAD method. However, when using the neural-network-based AAD method, the variability of the attended correlation is larger compared to using the least-squares-based AAD method. Fig. ? depicts the unattended correlation for different conditions. It can be observed that there is no significant difference in the unattended correlation obtained by either the neural-network-based AAD method or the least-squares-based AAD method. However, the variability of the unattended correlation is larger for the neural-network-based AAD method compared to using the least-squares-based AAD method. These results show that the neural-network-based AAD method is able to reconstruct the attended speech envelope with a better accuracy compared to the least-squares-based AAD method, but generates correlation coefficients with a larger variability.

For all acoustic conditions, Fig. \ref{fig: decoding performance} depicts the decoding performance when using either the least-squares-based AAD method, the neural-network-based AAD method, the state-space model with the least-squares-based AAD method, or the state-space model with neural-network-based AAD method. 
It can be observed that when using either the least-squares-based AAD method or the neural-network-based AAD method, a relatively low decoding performance (with the median decoding performance $69\%-73\%$) is obtained, mainly due to quite small (attended and unattended) correlations with a large variability (see Fig. \ref{fig: attended and unattended correlations}) based on which decoding is performed by these methods. A statistical multiple comparison test (Kruskal-Wallis test followed by the posthoc Dunn and Sidak test \cite{Hochberg1987}) revealed no significant difference ($p>0.05$) in decoding performance when using the least-squares-based AAD method or the neural-network-based AAD method. When using the state-space model with either the least-squares-based or the neural-network-based AAD method (LS--SSM, LS--NN), the decoding performance increases. The increase is considerably larger for the least-squares-based AAD method (with the median decoding performance $>94\%$) compared to the neural-network-based AAD method (with the median decoding performance $>73\%$). The larger decoding performance can be explained by the fact that the correlations generated by the least-squares-based AAD method have a lower variability compared to the correlations generated by the neural-network-based AAD method, which leads to a smoother estimate of attention probabilities and a more stable decoding. The statistical multiple comparison test revealed that for most acoustic conditions (except anechoic--noisy) the decoding performance using the state-space model with the least-squares-based AAD method is significantly larger ($p<0.05$) compared to using the least-squares-based AAD method, the neural-network-based AAD method, and the state-space model with the neural-network-based AAD method.

\section{Conclusion}
\label{sec:conclusion}
In this paper, we investigated a state-space model using correlation coefficients obtained with a 5-second correlation window to improve the decoding performance of the (linear) least-squares-based AAD method and the (non-linear) neural-network-based AAD method. 
The state-space model translates correlation coefficients, generated either by the least-squares-based or the neural-network-based AAD method, into smooth estimates of the attention state. 
The experimental results showed for all acoustic conditions that there is no significant difference in decoding performance between using the least-squares-based AAD method and the neural-network-based AAD method. However, when using the state-space model with the least-squares-based AAD method, for most acoustic conditions the decoding performance significantly improves. 

%the state-space model significantly improves the decoding performance, in particular when correlation coefficients are generated by the least-squares-based AAD method. In addition, the results showed that the least-squares-based AAD method and the neural-network-based AAD method yield a very similar decoding performance.

%(anechoic, reverberant, noisy, and reverberant-noisy). 

\bibliographystyle{IEEEtran}%IEEEbib, IEEEtran 
\bibliography{strings_2}

\begin{thebibliography}{10}
\providecommand{\url}[1]{#1}
\csname url@samestyle\endcsname
\providecommand{\newblock}{\relax}
\providecommand{\bibinfo}[2]{#2}
\providecommand{\BIBentrySTDinterwordspacing}{\spaceskip=0pt\relax}
\providecommand{\BIBentryALTinterwordstretchfactor}{4}
\providecommand{\BIBentryALTinterwordspacing}{\spaceskip=\fontdimen2\font plus
\BIBentryALTinterwordstretchfactor\fontdimen3\font minus
  \fontdimen4\font\relax}
\providecommand{\BIBforeignlanguage}[2]{{%
\expandafter\ifx\csname l@#1\endcsname\relax
\typeout{** WARNING: IEEEtran.bst: No hyphenation pattern has been}%
\typeout{** loaded for the language `#1'. Using the pattern for}%
\typeout{** the default language instead.}%
\else
\language=\csname l@#1\endcsname
\fi
#2}}
\providecommand{\BIBdecl}{\relax}
\BIBdecl

\bibitem{OSullivan_neural_2017}
J.~O'Sullivan, Z.~Chen, J.~Herrero, G.~M. McKhann, S.~A. Sheth, A.~D. Mehta,
  and N.~Mesgarani, ``Neural decoding of attentional selection in multi-speaker
  environments without access to clean sources,'' \emph{Journal of Neural
  Engineering}, vol.~14, no.~5, p. 56001, 2017.

\bibitem{Simon_Eyndhoven_2016}
S.~Van~Eyndhoven, T.~Francart, and A.~Bertrand, ``{EEG}-informed attended
  speaker extraction from recorded speech mixtures with application in
  neuro-steered hearing prostheses,'' \emph{IEEE Transactions on Biomedical
  Engineering}, vol.~64, no.~5, pp. 1045--1056, 2017.

\bibitem{Aroudi_2020_cognitive-driven_beamforming_ASLP}
A.~{Aroudi} and S.~{Doclo}, ``Cognitive-driven binaural beamforming using
  {EEG}-based auditory attention decoding,'' \emph{IEEE Transactions on Audio,
  Speech, and Language Processing}, in press.

\bibitem{OSullivan2014}
J.~A. O'Sullivan, A.~J. Power, N.~Mesgarani, S.~Rajaram, J.~J. Foxe, B.~G.
  Shinn-Cunningham, M.~Slaney, S.~A. Shamma, and E.~C. Lalor, ``Attentional
  selection in a cocktail party environment can be decoded from single-trial
  {{EEG}},'' \emph{Cerebral Cortex}, 2014.

\bibitem{Wong_Frontiers_2018}
\BIBentryALTinterwordspacing
D.~D. Wong, S.~A. Fuglsang, J.~Hjortkj{\ae}r, E.~Ceolini, M.~Slaney, and
  A.~de~Cheveign{\'e}, ``A comparison of regularization methods in forward and
  backward models for auditory attention decoding,'' \emph{Frontiers in
  Neuroscience}, vol.~12, p. 531, 2018. [Online]. Available:
  \url{https://www.frontiersin.org/article/10.3389/fnins.2018.00531}
\BIBentrySTDinterwordspacing

\bibitem{Alickovic_2019}
E.~Alickovic, T.~Lunner, F.~Gustafsson, and L.~Ljung, ``A tutorial on auditory
  attention identification methods,'' \emph{Frontiers in Neuroscience},
  vol.~13, p. 153, 2019.

\bibitem{Sina_2018}
S.~Miran, S.~Akram, A.~Sheikhattar, J.~Z. Simon, T.~Zhang, and B.~Babadi,
  ``Real-time tracking of selective auditory attention from {M/EEG}: A
  {Bayesian} filtering approach,'' \emph{Frontiers in Neuroscience}, vol.~12,
  p. 262, 2018.

\bibitem{Tobias_2018}
T.~de~Taillez, B.~Kollmeier, and B.~T. Meyer, ``Machine learning for decoding
  listeners� attention from electroencephalography evoked by continuous
  speech,'' \emph{European Journal of Neuroscience}, Dec. 2018.

\bibitem{Gregory_2019}
G.~Ciccarelli, M.~Nolan, J.~Perricone, P.~T. Calamia, S.~Haro, J.~O’Sullivan,
  N.~Mesgarani, T.~F. Quatieri, and C.~J. Smalt, ``Comparison of two-talker
  attention decoding from {EEG} with nonlinear neural networks and linear
  methods,'' \emph{Scientific Reports, Nature}, vol.~9, no. 11538, Aug. 2019.

\bibitem{Aroudi_ICASSP2019}
A.~Aroudi and S.~Doclo, ``Cognitive-driven binaural {LCMV} beamformer using
  {EEG}-based auditory attention decoding,'' in \emph{Proc. IEEE International
  Conference on Acoustics, Speech and Signal Processing {(ICASSP)}}, Brighton,
  United Kingdom, May 2019, pp. 406--410.

\bibitem{Cong_2019}
C.~Han, J.~O{\textquoteright}Sullivan, Y.~Luo, J.~Herrero, A.~D. Mehta, and
  N.~Mesgarani, ``Speaker-independent auditory attention decoding without
  access to clean speech sources,'' \emph{Science Advances}, vol.~5, no.~5,
  2019.

\bibitem{Sahar_2016}
S.~Akram, J.~Z. Simon, and B.~Babadi, ``Dynamic estimation of the auditory
  temporal response function from {MEG} in competing-speaker environments,''
  \emph{IEEE Transactions on Biomedical Engineering}, vol.~64, no.~8, pp.
  1896--1905, 2017.

\bibitem{Aroudi_EMBC_2017}
A.~Aroudi and S.~Doclo, ``{EEG-based} auditory attention decoding using
  unprocessed binaural signals in reverberant and noisy conditions,'' in
  \emph{Proc. Int. Conf. of the IEEE Engineering in Medicine and Biology
  Society (EMBC)}, Jeju, South Korea, 2017, pp. 484--488.

\bibitem{Kayser2009}
H.~Kayser, S.~D. Ewert, J.~Anem{\"u}ller, T.~Rohdenburg, V.~Hohmann, and
  B.~Kollmeier, ``Database of multichannel in-ear and behind-the-ear
  head-related and binaural room impulse responses,'' \emph{EURASIP Journal on
  Advances in Signal Processing}, vol. 2009, p.~6, 2009.

\bibitem{Habets2008}
E.~Habets, I.~Cohen, and S.~Gannot, ``Generating nonstationary multisensor
  signals under a spatial coherence constraint,'' \emph{Journal of the
  Acoustical Society of America}, vol. 124, no.~5, pp. 2911--2917, Nov. 2008.

\bibitem{Ali_2019_neural_system}
A.~{Aroudi}, B.~{Mirkovic}, M.~{De Vos}, and S.~{Doclo}, ``Impact of different
  acoustic components on {EEG}-based auditory attention decoding in noisy and
  reverberant conditions,'' \emph{IEEE Transactions on Neural Systems and
  Rehabilitation Engineering}, vol.~27, no.~4, pp. 652--663, April 2019.

\bibitem{Dozat2016}
T.~Dozat, ``Incorporating nesterov momentum into adam,'' in \emph{International
  Conference on Learning Representations {(ICLR 2016 workshop)}}, 2016.

\bibitem{Srivastava_2014}
N.~Srivastava, G.~Hinton, A.~Krizhevsky, I.~Sutskever, and R.~Salakhutdinov,
  ``Dropout: A simple way to prevent neural networks from overfitting,''
  \emph{Journal of Machine Learning Research}, vol.~15, pp. 1929--1958, 2014.

\bibitem{chollet2015keras}
F.~Chollet \emph{et~al.}, ``Keras,'' \url{https://keras.io}, 2015.

\bibitem{Hochberg1987}
Y.~Hochberg and A.~C. Tamhane, \emph{Multiple Comparison Procedures}.\hskip 1em
  plus 0.5em minus 0.4em\relax John Wiley and Sons, 1987.

\end{thebibliography}

\end{document}